\documentclass[10pt, conference, a4paper]{IEEEtran}
\usepackage{amsmath, amsfonts,  amssymb, color, textcomp}
\usepackage[all]{xy}
\usepackage[dvips]{graphicx}
\usepackage{latexsym}
\usepackage{epsfig}
\usepackage{bm}
\usepackage{xspace}
\usepackage{url}
\newcommand{\ms}{m_{\text{s}}}

\newcommand{\code}[1]{\mathsf{#1}}

\newcommand{\codeCQC}[1]{\code{C}_{\mathrm{QC}}^{(r)}}

\renewcommand{\baselinestretch}{1.0}

\begin{document}

\title{Asymptotically Good LDPC Convolutional Codes Based on Protographs}
\author{
\authorblockN{David G. M. Mitchell$^*$, Ali E. Pusane$^\dag$, Kamil Sh. Zigangirov$^\dag$, and Daniel J. Costello, Jr.$^\dag$}
\authorblockA{$^*$Institute for Digital Communications, Joint Research Institute for Signal \& Image
Processing,\\
The University of Edinburgh, Scotland,
David.Mitchell@ed.ac.uk\\
$^\dag$Dept. of Electrical Engineering, University of Notre Dame,
Notre Dame,
Indiana, USA,\\
\{apusane, kzigangi, dcostel$1$\}@nd.edu}}

\maketitle

\begin{abstract}
LDPC convolutional codes have been shown to be capable of achieving
the same capacity-approaching performance as LDPC block codes with
iterative message-passing decoding. In this paper, asymptotic
methods are used to calculate a lower bound on the free distance for
several ensembles of asymptotically good protograph-based LDPC convolutional codes.
Further, we show that the free distance to constraint length ratio
of the LDPC convolutional codes exceeds the minimum distance to
block length ratio of corresponding LDPC block codes.
\end{abstract}

\section{Introduction}\label{sec:intro}
Along with turbo codes, low-density parity-check (LDPC) block codes
form a class of codes which approach the (theoretical) Shannon
limit. LDPC codes were first introduced in the $1960$s by Gallager
$\cite{gal}$. However, they were considered impractical at that time and very little related work was done until Tanner provided a graphical interpretation of the parity-check matrix in 1981 $\cite{tan}$. More recently, in his Ph.D. Thesis, Wiberg revived interest in LDPC codes and further developed the relation between
Tanner graphs and iterative decoding \cite{wib}.

The convolutional counterpart of LDPC block codes was introduced in
\cite{fels}, and LDPC convolutional codes have been shown to have
certain advantages compared to LDPC block codes of the same
complexity \cite{cost,cost2}. In this paper, we use ensembles of tail-biting LDPC
convolutional codes derived from a protograph-based ensemble of LDPC
block codes to obtain a lower bound on the free distance of
unterminated, asymptotically good, periodically time-varying LDPC convolutional code ensembles,
i.e., ensembles that have the property of free distance growing linearly with
constraint length.

In the process, we show that the minimum distances of ensembles
of tail-biting LDPC convolutional codes (introduced in
\cite{zigvde}) approach the free distance of an associated
unterminated, periodically time-varying LDPC convolutional code
ensemble as the block length of the tail-biting ensemble increases.
We also show that, for protographs with regular degree
distributions, the free distance bounds are consistent with those
recently derived for regular LDPC convolutional code ensembles in
\cite{srid} and \cite{truh}. Further, for protographs with irregular
degree distributions, we obtain new free distance bounds that grow
linearly with constraint length and whose free distance to
constraint length ratio exceeds the minimum distance to block length
ratio of the corresponding block codes.

The paper is structured as follows. In Section \ref{sec:ldpccc}, we
briefly introduce LDPC convolutional codes. Section \ref{sec:proto}
summarizes the technique proposed by Divsalar to analyze the
asymptotic distance growth behavior of protograph-based LDPC block
codes \cite{div}. In Section \ref{sec:distbnd}, we describe the
construction of tail-biting LDPC convolutional codes as well as the
corresponding unterminated, periodically time-varying LDPC
convolutional codes. We then show that the free distance of a
periodically time-varying LDPC convolutional code is lower bounded
by the minimum distance of the block code formed by terminating it
as a tail-biting LDPC convolutional code. Finally, in Section
\ref{sec:results} we present new results on the free distance of
ensembles of LDPC convolutional codes based on protographs.

\section{LDPC convolutional codes}\label{sec:ldpccc}
We start with a brief definition of a rate $R=b/c$ binary LDPC
convolutional code $\mathcal{C}$. (A more detailed description can
be found in \cite{fels}.) A code sequence $\mathbf{v}_{[0,\infty ]}$
satisfies the equation
\begin{equation}\label{codeword1}
\mathbf{v}_{[0,\infty ]}\mathbf{H}_{[0,\infty
]}^{\mathtt{T}}=\mathbf{0},
\end{equation}
where $\mathbf{H}_{[0,\infty]}^{\mathtt{T}}$ is the syndrome former matrix and

\noindent $\mathbf{H}_{[0,\infty]}=$ \vspace{1mm}\\
\scalebox{0.96}{\mbox{\scriptsize{$
\left[ \begin{array}{cccccc}
\mathbf{H}_{0}(0) & & & & \\
\mathbf{H}_{1}(1) & \mathbf{H}_{0}(1)& & & \\
\vdots & \vdots& & \ddots& \\
\mathbf{H}_{\ms}(\ms) & \mathbf{H}_{\ms-1}(\ms)&\ldots &\mathbf{H}_{0}(\ms) & \\
& \mathbf{H}_{\ms}(\ms+1) & \mathbf{H}_{\ms-1}(\ms+1)&\ldots
&\mathbf{H}_{0}(\ms+1) \\
&\ddots&\ddots&&\ddots
\end{array} \right]$}
}}

\noindent is the parity-check matrix of the convolutional code
$\mathcal{C}$. The submatrices $\mathbf{H}_{i}(t)$,
$i=0,1,\cdots,\ms$, $t\geq 0$,  are binary $\left( c-b\right) \times c$
submatrices, given by
\begin{equation}
\mathbf{H}_i(t)= \left[ \begin{array}{ccc}
h_i^{(1,1)}(t) & \cdots & h_i^{(1,c)}(t) \\
\vdots & & \vdots \\
h_i^{(c-b,1)}(t) & \cdots & h_i^{(c-b,c)}(t) \\
\end{array} \right],\end{equation}
that satisfy the following properties:
\begin{enumerate}
\item $\mathbf{H}_{i}(t)=\mathbf{0},~i<0$ and $i>m_{s},~\forall ~t.$
\item There is a $t$ such that $\mathbf{H}_{\ms}(t) \neq \mathbf{0}.$
\item $\mathbf{H}_{0}(t)\neq\mathbf{0}$ and has full rank $\forall \,\, t$.
\end{enumerate}
We call $\ms$ the syndrome former memory and $\nu_{\text{s}} =
(\ms+1)\cdot c$ the decoding constraint length. These parameters
determine the width of the
nonzero diagonal region of $\mathbf{H}_{[0,\infty]}.$ %
The sparsity of the parity-check matrix is ensured by demanding that its rows have very low Hamming weight, i.e., %
$w_{H}(\mathbf{h}_{i})<<(\ms+1)\cdot c,~i>0$, where $\mathbf{h}_{i}$ denotes the $i$-th row of %
$\mathbf{H}_{[0,\infty ]}$. The code is said to be regular if its
parity-check matrix $\mathbf{H}_{[0,\infty]}$ has exactly $J$ ones
in every column and, starting from row $(c-b)\ms+1$, $K$ ones in every row. The other entries are
zeros. We refer to a code with these properties as an
$(m_\text{s},J,K)$-regular LDPC convolutional code, and we note that,
in general, the code is time-varying and has rate $R=1-J/K$. An
$(m_\text{s},J,K)$-regular time-varying LDPC convolutional code is
periodic with period $T$ if $\mathbf{H}_{i}(t)$ is periodic, i.e.,
$\mathbf{H}_{i}(t)=\mathbf{H}_{i}(t+T), \forall ~i,t$, and if
$\mathbf{H}_{i}(t)=\mathbf{H}_{i}, \forall ~i,t$, the code is
time-invariant.

An LDPC convolutional code is called irregular if its row and column
weights are not constant. The notion of degree distribution is used
to characterize the variations of check and variable node degrees in
the Tanner graph corresponding to an LDPC convolutional code.
Optimized degree distributions have been used to design LDPC
convolutional codes with good iterative decoding performance in the
literature (see, e.g., \cite{zigvde,richter,arv,pus}), but no
distance bounds for irregular LDPC convolutional code ensembles have
been previously published.

\section{Protograph Weight Enumerators}\label{sec:proto}
\begin{figure}[htp]\begin{center}
\includegraphics[width=3.5in]{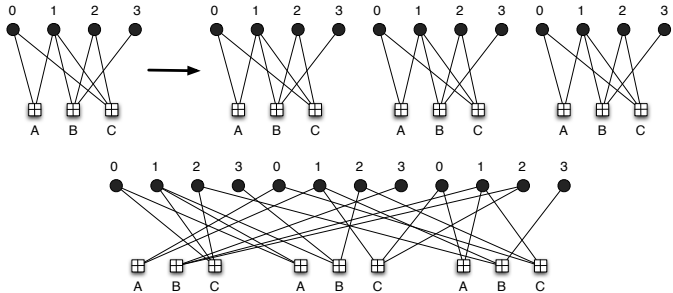}
\end{center}
\caption{The copy-and-permute operation for a
protograph.}\label{fig:proto}
\end{figure}

Suppose a given protograph has $n_v$ variable nodes and $n_c$ check nodes. An
ensemble of protograph-based LDPC block codes can be created by the copy-and-permute
operation \cite{thor}. The Tanner graph obtained for one member of
an ensemble created using this method is illustrated in Fig.
\ref{fig:proto}.
\vspace{-1ex}
\vspace{-1ex}

The parity-check matrix $H$ corresponding to the ensemble of
protograph-based LDPC block codes can be obtained by replacing ones
with $N \times N$ permutation matrices and zeros with $N \times N$
all zero matrices in the underlying protograph parity-check matrix
$P$, where the permutation matrices are chosen randomly and
independently. The protograph parity-check matrix $P$ corresponding
to the protograph given in Figure \ref{fig:proto} can be written as
\vspace{-1.5ex}
\begin{center}
\includegraphics[width=1.2in]{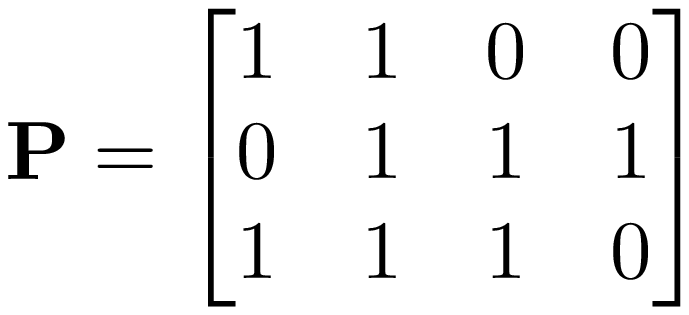}\raisebox{6mm}{,}
\end{center}
\vspace{-2ex}
where we note that, since the row and column weights of $P$ are not
constant, $P$ represents the parity-check matrix of an irregular
LDPC code. If a variable node and a check node in the protograph are
connected by $r$ parallel edges, then the associated entry in $P$
equals $r$ and the corresponding block of $H$ consists of a
summation of $r$ $N \times N$ permutation matrices. The sparsity
condition of an LDPC parity-check matrix is thus satisfied for large
$N$. The code created by applying the copy-and-permute operation to
an $n_c\times n_v$ protograph parity-check matrix $P$ has block
length $n=Nn_v$. In addition, the code has the same rate and degree
distribution for each of its variable and check nodes as the
underlying protograph code.

Combinatorial methods of calculating ensemble average weight
enumerators have been presented in \cite{div} and \cite{fog}. The remainder of
this Section summarizes the methods presented in \cite{div}.
\subsection{Ensemble weight enumerators}

Suppose a protograph contains $m$ variable nodes to be transmitted
over the channel and $n_v-m$ punctured variable nodes. Also, suppose
that each of the $m$ transmitted variable nodes has an associated
weight $d_i$, where $0 \leq d_i \leq N$ for all $i$.\footnote{Since
we use $N$ copies of the protograph, the weight associated with a
particular variable node in the protograph can be as large as $N$.}
Let $S_d =\{(d_1,d_2,\ldots,d_m)\}$ be the set of all possible
weight distributions such that $d_1+\ldots+d_m=d$, and let $S_p$ be
the set of all possible weight distributions for the remaining
punctured nodes. The ensemble weight enumerator for the protograph
is then given by \vspace{-1ex}
\begin{equation}
A_d = \sum_{\{d_k\}\in S_d}\sum_{\{d_j\}\in S_p}A_\mathbf{d},
\end{equation}
where $A_\mathbf{d}$ is the average number of codewords in the
ensemble with a particular weight distribution
$\textbf{d}=(d_1,d_2,\ldots,d_{n_v})$.
\subsection{Asymptotic weight enumerators}
The normalized logarithmic asymptotic weight distribution of a code
ensemble can be written as $r(\delta) = \lim_{n\rightarrow
\infty}\textrm{sup } r_n(\delta),$ where $r_n(\delta)
=\frac{\textrm{ln}(A_d)}{n}$, $\delta = d/n$, $d$ is the Hamming
distance, $n$ is the block length, and $A_d$ is the ensemble average
weight distribution.

Suppose the first zero crossing of $r(\delta)$ occurs at $\delta =
\delta_{min}$.  If $r(\delta)$ is negative in the range $
0<\delta<\delta_{min}$, then $\delta_{min}$ is called
the \emph{minimum distance growth rate} of the code ensemble. By
considering the probability\vspace{-1ex}
$$\mathbb{P}(d < \delta_{min} n) =
\sum^{\delta_{min}n-1}_{d=1}A_d,$$

\vspace{-1ex}

\noindent it is clear that, as the
block length $n$ grows, if $\mathbb{P}(d < \delta_{min} n)
<<1$, then we can say with high probability that the majority of
codes in the ensemble have a minimum distance that grows linearly
with $n$ and that the distance growth rate is $\delta_{min}$.
\section{Free distance bounds} \label{sec:distbnd} In this section we
present a method for obtaining a lower bound on the free distance of
an ensemble of unterminated, asymptotically good, periodically
time-varying LDPC convolutional codes derived from protograph-based
LDPC block codes. To proceed, we will make use of a family of
tail-biting LDPC convolutional codes with incremental increases in
block length. The tail-biting codes will be used as a tool to obtain
the desired bound on the free distance of the unterminated codes.

\subsection{Tail-biting convolutional codes} \label{sec:tb}
Suppose that we have an $n_c\times n_v$ protograph parity-check
matrix $P$, where gcd$(n_c,n_v) = y$. We then partition $P$ as a $y
\times y$ block matrix as follows:
$$ P = \left[\begin{array}{ccc}
P_{1,1} &\ldots&P_{1,y}\\
\vdots&&\vdots\\
P_{y,1}&\ldots&P_{y,y}
\end{array}\right],$$
\noindent where each block $P_{i,j}$ is of size $n_c/y \times
n_v/y$. $P$ can thus be separated into a lower triangular part,
$P_l$, and an upper triangular part minus the leading diagonal,
$P_u$. Explicitly,

\vspace{2mm}\noindent\hspace{-1mm} \scalebox{0.95}{\mbox{\scriptsize{$P_l =
\left[
\begin{array}{cccc}
P_{1,1} &&\\
P_{2,1}&P_{2,2}&&\\
\vdots&\vdots&\ddots&\\
P_{y,1}&P_{y,2}&\ldots&P_{y,y}
\end{array}
\right] \textrm{ and } P_u = \left[\begin{array}{cccc}
&P_{1,2}&\ldots&P_{1,y}\\
&&\ddots&\vdots\\
&&&P_{y-1,y}\\
&&&
\end{array}\right]$,}
}}\vspace{2mm}

\noindent where blank spaces correspond to zeros. This operation is
called `cutting' a protograph parity-check matrix.

Rearranging the positions of these two triangular matrices and
repeating them indefinitely results in a parity-check matrix
$H_{cc}$ of an unterminated, periodically time-varying convolutional
code with constraint length $\nu_{\text{s}}=n_v$ and period $T=y$
given by\footnote{Cutting certain protograph parity-check matrices
may result in a smaller period $T=y^{\prime}$ of $H_{\textrm{cc}}$,
where $y^{\prime} \in \mathbb{Z}^+$ divides $y$ without remainder.
If $y^{\prime}=1$ then the resulting convolutional code is
time-invariant.}
\begin{equation}
H_{\textrm{cc}} = \left[\begin{array}{cccc}
P_l &&&\\
P_u&P_l&&\\
&P_u&P_l&\\
&&\ddots&\ddots
\end{array}\right].
\end{equation}

\noindent Note that if gcd$(n_c,n_v) = 1$, we cannot form a square
block matrix larger than $1 \times 1$ with equal size blocks. In
this case, $P_l = P$ and $P_u$ is the all zero matrix of size $n_c
\times n_v$. This trivial cut results in a convolutional code with
syndrome former memory zero, with repeating blocks of the original
protograph on the leading diagonal. It is necessary in this case to
create a larger protograph parity-check matrix by using the copy and
permute operation on $P$. This results in an $M n_c \times M n_v =
n_c^\prime \times n_v^\prime$ parity-check matrix for some small
integer $M$. The $n_c^\prime \times n_v^\prime$ protograph
parity-check matrix can then be cut following the procedure outlined
above. In effect, the choice of $M \times M$ permutation matrix
creates a mini ensemble of block codes suitable to be unwrapped to
an ensemble of convolutional codes.

We now introduce the notion of tail-biting convolutional codes by
defining an `unwrapping factor' $\lambda$ as the number of times the
sliding convolutional structure is repeated. For $\lambda > 1$, the
parity-check matrix $H_{\textrm{tb}}^{(\lambda)}$ of the desired
tail-biting convolutional code can be written as
$$H_{\textrm{tb}}^{(\lambda)} = \left[\begin{array}{ccccc}
P_l &&&&P_u\\
P_u&P_l&&&\\
&P_u&P_l&&\\
&&\ddots&\ddots&\\
&&&P_u&P_l
\end{array}\right]_{\lambda n_c \times \lambda n_v}.$$
\vspace{-2mm}

\noindent Note that the tail-biting convolutional code for $\lambda
= 1$ is simply the original block code.

\subsection{A tail-biting LDPC convolutional code ensemble}\label{sec:tailll}
Given a protograph parity-check matrix $P$, we generate a family of
tail-biting convolutional codes with increasing block lengths
$\lambda n_v$, $\lambda = 1,2,\ldots$, using the process described
above. Since tail-biting convolutional codes are themselves block
codes, we can treat the Tanner graph of
$H_{\textrm{tb}}^{(\lambda)}$ as a protograph for each value of
$\lambda$. Replacing the entries of this matrix with either $N
\times N$ permutation matrices or $N \times N$ all zero matrices, as
discussed in Section \ref{sec:proto}, creates an ensemble of LDPC
codes that can be analyzed asymptotically as $N$ goes to infinity,
where the sparsity condition of an LDPC code is satisfied for large
$N$. Each tail-biting LDPC code ensemble, in turn, can be unwrapped
and repeated indefinitely to form an ensemble of unterminated,
periodically time-varying LDPC convolutional codes with constraint
length $\nu_s = N n_v$ and, in general, period $T=\lambda y$.

Intuitively, as $\lambda$ increases, the tail-biting code becomes a
better representation of the associated unterminated convolutional
code, with $\lambda \rightarrow \infty$ corresponding to the
unterminated convolutional code itself. This is reflected in the
weight enumerators, and it is shown in Section \ref{sec:results}
that increasing $\lambda$ provides us with distance growth rates
that converge to a lower bound on the free distance growth
rate of the unterminated convolutional code.

\subsection{A free distance bound}
Tail-biting convolutional codes can be used to establish a lower
bound on the free distance of an associated unterminated,
periodically time-varying convolutional code by showing that the
free distance of the unterminated code is lower bounded by the
minimum distance of any of its tail-biting versions. A proof can be
found in \cite{truh}.
\newtheorem{freedist}{Theorem}
\begin{freedist}
Consider a rate $R=(n_v-n_c)/n_v$ unterminated, periodically
time-varying convolutional code with decoding constraint length
$\nu_\text{s}=Nn_v$ and period $T=\lambda y$. Let $d_{min}$ be
the minimum distance of the associated tail-biting convolutional
code with length $n=\lambda N n_v$ and unwrapping factor
$\lambda>0$. Then the free distance $d_{free}$ of the unterminated
convolutional code is lower bounded by $d_{min}$ for any unwrapping
factor $\lambda$, i.e.,
\vspace{-2.5mm}
\begin{equation}\label{emre}
    d_{free} \geq d_{min} ,\quad \forall \lambda > 0.
\end{equation}
\end{freedist}

\vspace{1mm}
A trivial corollary of the above theorem is that the minimum
distance of a protograph-based LDPC block code is a lower bound on
the free distance of the associated unterminated, periodically
time-varying LDPC convolutional code. This can be observed by
setting $\lambda = 1$.

\subsection{The free distance growth rate}

One must be careful in comparing the distance growth rates of codes
with different underlying structures. A fair basis for comparison
generally requires equating the complexity of encoding and/or
decoding of the two codes. Traditionally, the minimum distance
growth rate of block codes is measured relative to block length,
whereas constraint length is used to measure the free distance
growth rate of convolutional codes. These measures are based on the
complexity of decoding both types of codes on a trellis. Indeed, the
typical number of states required to decode a block code on a
trellis is exponential in the block length, and similarly the number
of states required to decode a convolutional code is exponential in
the constraint length. This has been an accepted basis of comparing
block and convolutional codes for decades, since maximum-likelihood
decoding can be implemented on a trellis for both types of codes.

The definition of decoding complexity is different, however, for
LDPC codes. The sparsity of their parity-check matrices, along with
the iterative message-passing decoding algorithm typically employed,
implies that the decoding complexity per symbol depends on the
degree distribution of the variable and check nodes and is
independent of both the block length and the constraint length. The
cutting technique we described in Section \ref{sec:tb} preserves the
degree distribution of the underlying LDPC block code, and thus the
decoding complexity per symbol is the same for the block and
convolutional codes considered in this paper.

Also, for randomly constructed LDPC block codes, state-of-the-art
encoding algorithms require only $O(g)$ operations per symbol, where
$g<<n$ \cite{richardson}, whereas for LDPC convolutional codes, if
the parity-check matrix satisfies the conditions listed in Section
II, the number of encoding operations per symbol is only $O(1)$
\cite{pusaneTCOM}. Here again, the encoding complexity per symbol is
essentially independent of both the block length and the constraint
length.

Hence, to compare the distance growth rates of LDPC block and
convolutional codes, we consider the hardware complexity of
implementing the encoding and decoding operations in hardware.
Typical hardware storage requirements for both LDPC block encoders
and decoders are proportional to the block length $n$. The
corresponding hardware storage requirements for LDPC convolutional
encoders and decoders are proportional to the decoding constraint
length \cite{pusaneTCOM}.\footnote{For rates other than $1/2$,
encoding constraint lengths may be preferred to decoding constriant
lengths. For further details, see \cite{tc08}.}

\section{Results and Discussion}\label{sec:results}
\subsection{Distance growth rate results}
We now present distance growth rate results for several ensembles of rate $1/2$
asymptotically good LDPC convolutional codes based on protographs.

\begin{figure}[htp]\begin{center}
\includegraphics[width=3.2in]{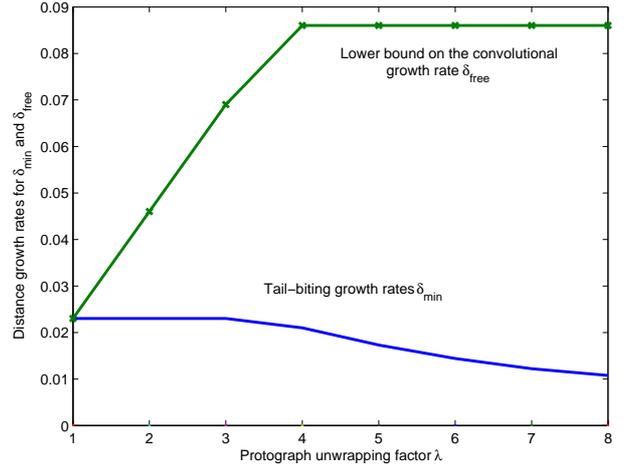}
\end{center}
\vspace{-2mm}
\caption{Distance growth rates for Example $1$.}\label{fig:ex1}
\end{figure}

\emph{Example $1$} Consider a $(3,6)$ regular LDPC code with the
folowing protograph:\vspace{-1.8mm}
\begin{center}
\includegraphics[width=1.5in]{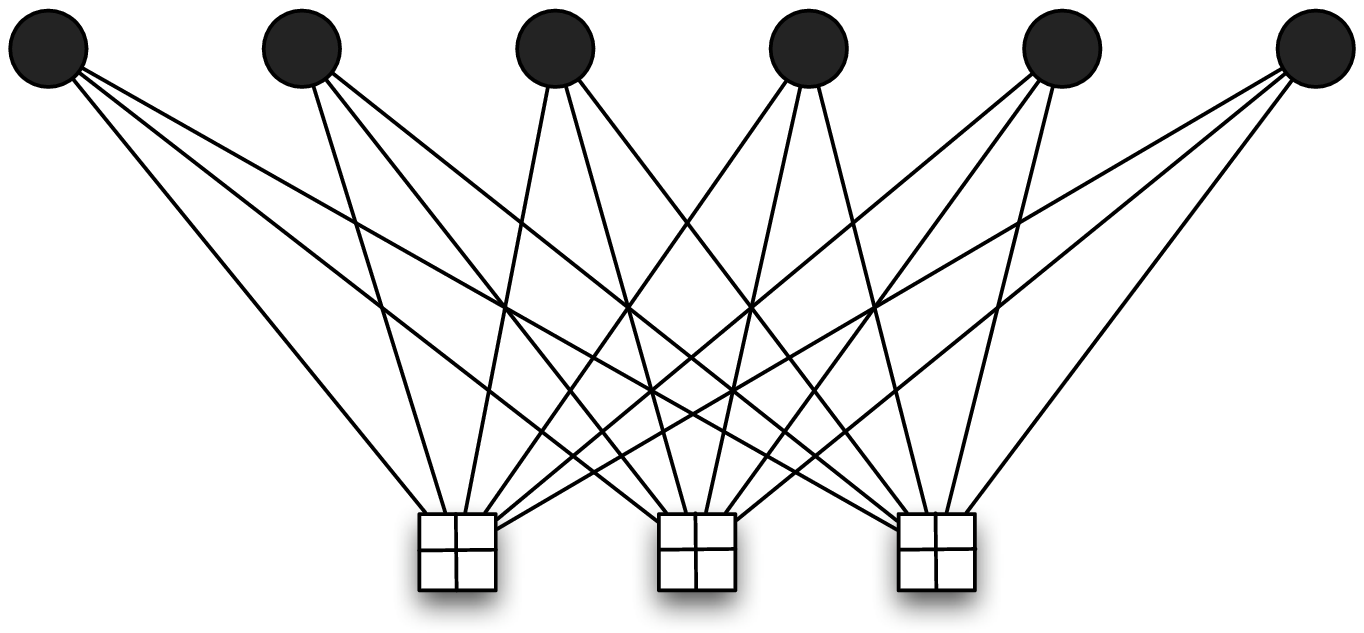}\raisebox{1cm}{.}
\end{center}
\vspace{-3.6mm}
\noindent For this example, the minimum distance growth rate is
$\delta_{min} = 0.023$, as originally calculated by Gallager \cite{gal}.
A family of tail-biting LDPC convolutional code ensembles can be
generated according to the following cut:\vspace{-1.7mm}
\begin{center}
\includegraphics[width=1.5in]{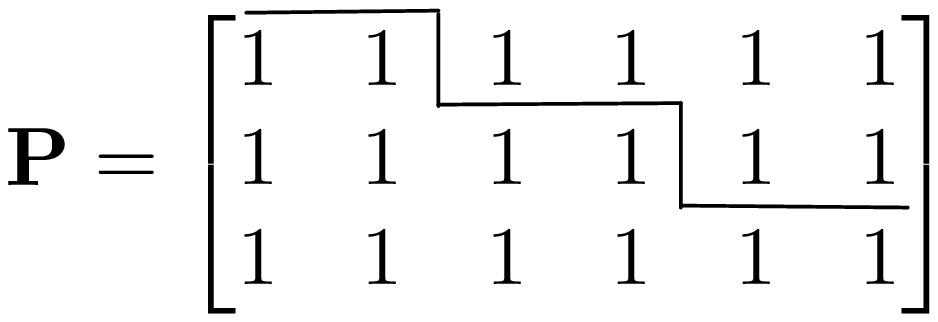}\raisebox{6mm}{.}
\end{center}
\vspace{-3.5mm}
For each $\lambda$, the minimum distance growth rate
$\delta_{min}$ was calculated for the tail-biting LDPC
convolutional codes using the approach outlined in Section
\ref{sec:tailll}. The distance growth rates for each $\lambda$ are given as
\begin{equation}
\delta_{min} = \frac{d_{min}}{n}=\frac{d_{min}}{\lambda N
n_v}=\frac{d_{min}}{\lambda \nu_\text{s}}.
\end{equation}
The free distance growth rate of the associated rate $1/2$ ensemble of
unterminated, periodically time-varying LDPC convolutional codes is
$\delta_{free}=d_{free}/\nu_\text{s}$, as discussed above. Then (\ref{emre})
gives us the lower bound
\begin{equation}\label{lb}
 \delta_{free}=\frac{d_{free}}{\nu_\text{s}}\geq
\frac{d_{min}}{\nu_\text{s}}=\lambda
\delta_{min}
\end{equation}
for $\lambda\geq 1$. These growth rates are plotted in Fig. \ref{fig:ex1}.
\vspace{-3mm}
\vspace{-3mm}

We observe that, once the unwrapping factor $\lambda$ of the
tail-biting convolutional codes exceeds $3$, the lower bound on
$\delta_{free}$ levels off at $\delta_{free}\geq 0.086$, which agrees with the
results
presented in \cite{srid} and \cite{truh} and represents a significant increase
over the value of $\delta_{min}$. In this case, the minimum
weight codeword in the unterminated convolutional code also appears
as a codeword in the tail-biting code.

\emph{Example $2$} The following irregular protograph is from the
Repeat Jagged Accumulate \cite{divdo} (RJA) family. It was shown to
have a good iterative decoding threshold ($\gamma_\textrm{iter} =
1.0$ dB) while maintaining linear minimum distance growth
($\delta_{min}=0.013$). We display below the associated $P$
matrix and cut used to generate the family of tail-biting LDPC
convolutional code ensembles.
\vspace{-1.8mm}
\begin{center}
\includegraphics[width=1.2in]{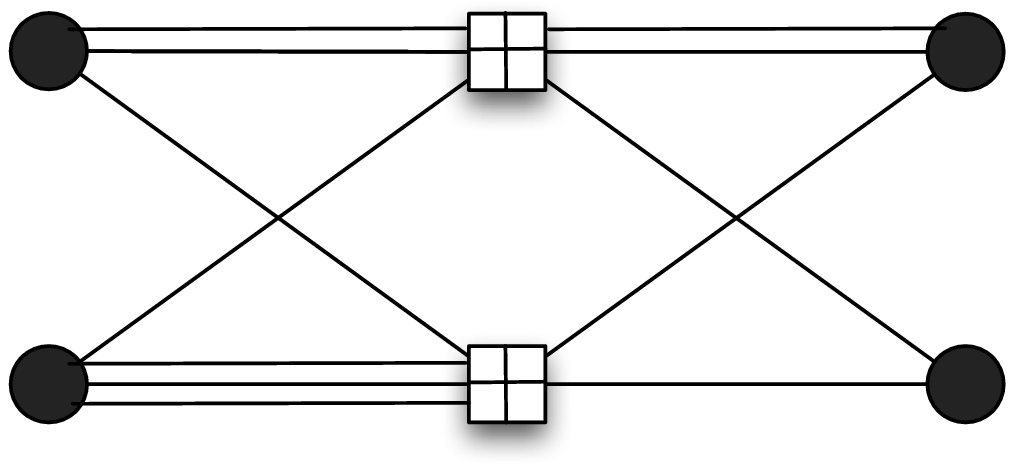}\hspace{2mm}\raisebox{7mm}{
$\leftrightsquigarrow$ }
\raisebox{2mm}{ \includegraphics [ width=1.2in]{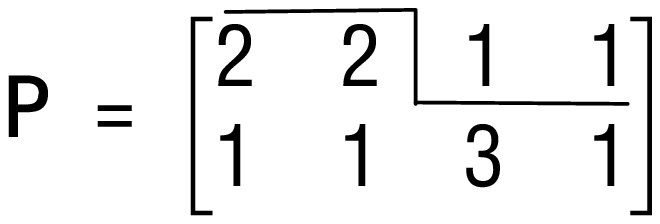}}\raisebox{7mm}{.}
\end{center}
\vspace{-4mm}
\noindent  We observe that, as in Example $1$, the minimum distance
growth rates calculated for increasing $\lambda$ provide us with a
lower bound on the free distance growth rate of the convolutional code
ensemble using (\ref{lb}). The lower bound was calculated as $\delta_{free}\geq
0.057$ (for $\lambda
\geq 5$), significantly larger than the minimum distance
growth rate $\delta_{min}$ of the underlying block code
ensemble.

\emph{Example $3$} The following irregular protograph is from the Accumulate
Repeat Jagged Accumulate family (ARJA) \cite{divdo}:
\vspace{-1.8mm}
\begin{center}
\includegraphics[width=2.1in]{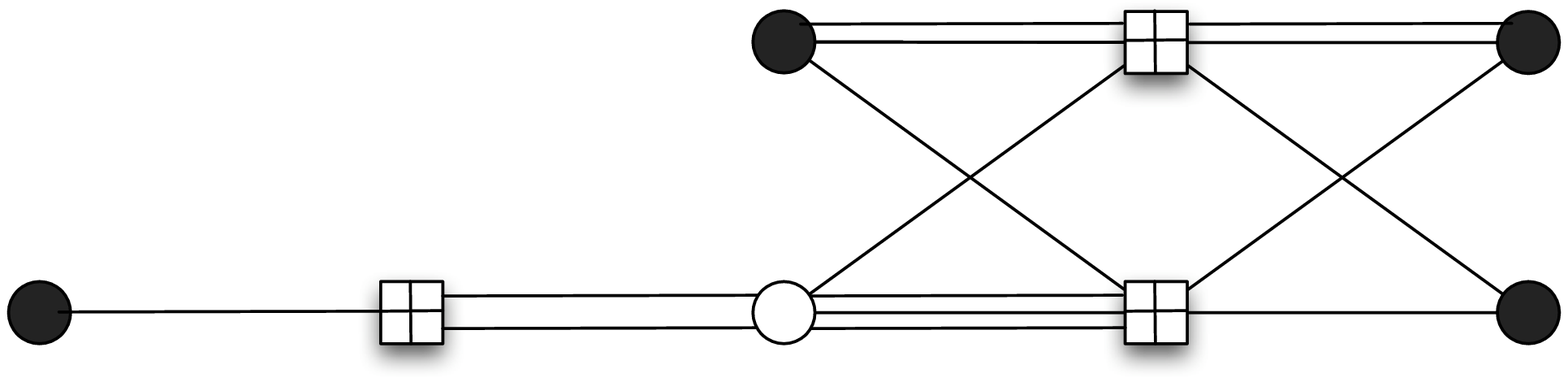}\raisebox{7mm}{,}
\end{center}
\vspace{-3.4mm}

\noindent where the undarkened circle represents a punctured
variable node. This protograph is of significant practical interest,
since it was shown to have $\delta_{min}=0.015$ and iterative
decoding threshold $\gamma_\textrm{iter}=0.628$, i.e., pre-coding
the protograph of Example $2$ provides an improvement in both
values.

In this ARJA example, the protograph matrix $P$ is of size $n_c
\times n_v = 3 \times 5$. We observe that gcd$(n_c,n_v) = 1$, and
thus we have the trivial cut mentioned in Section \ref{sec:tb}. We
must then copy and permute $P$ to generate a mini ensemble of block
codes. Results are shown for one particular member of the mini
ensemble with $M=2$, but a change in performance can be obtained by
varying the particular permutation chosen. Increasing $\lambda$ for
the chosen permutation results in a lower bound, found using (\ref{lb}),
of $\delta_{free} \geq 0.053$ for $\lambda \geq 4$. Again, we
observe a significant increase in $\delta_{free}$ compared to
$\delta_{min}$. \vspace{-1ex}
\subsection{Simulation results}
\begin{figure}[htp]\begin{center}
\includegraphics[width=3.5in]{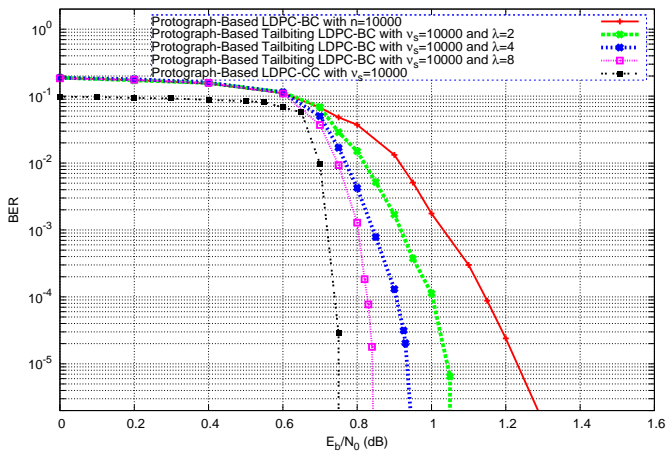}
\end{center}
\vspace{-3ex} \caption{Simulation results for Example
$3$.}\label{fig:sim}
\end{figure}

\vspace{-0.5ex}
Simulation results for LDPC block and convolutional codes based on
the protograph of Example $3$ were obtained assuming BPSK modulation
and an additive white Gaussian noise channel (AWGNC). All decoders
were allowed a maximum of $100$ iterations, and the block code
decoders employed a syndrome-check based stopping rule. As a result of their
block structure, tail-biting LDPC convolutional codes were decoded using
standard LDPC block decoders employing a belief-propagation decoding algorithm.
The LDPC convolutional code, on the other hand, was decoded by a sliding-window
based belief-propagation decoder \cite{fels}. The
resulting bit error rate (BER) performance is shown in
Fig.\ref{fig:sim}.

\vspace{-3ex}
\vspace{-2ex}

We note that the protograph-based tail-biting LDPC convolutional
codes outperform the underlying protograph-based LDPC block code
(which can also be seen as a tail-biting code with unwrapping factor
$\lambda=1$). Larger unwrapping factors yield improved error
performance, eventually approaching the performance of the
unterminated convolutional code, which can be seen as a tail-biting
code with an infinitely large unwrapping factor. We also note that
no error floor is observed for the convolutional code, which is
expected, since the code ensemble is asymptotically good and has a
relatively large ($\delta_{free} \geq 0.053$) distance
growth rate.

We also note that the performance of the unterminated LDPC
convolutional code is consistent with the iterative decoding
threshold computed for the underlying protograph. At a moderate
constraint length of $10000$, the unterminated code achieves
$10^{-5}$ BER at roughly $0.12$ dB away from the threshold, and with
larger block (constraint) lengths, the performance will improve even
further. This is expected, since both the unterminated and the
tail-biting convolutional codes preserve the same degree
distribution as the underlying protograph. \vspace{-1.5ex}

\section{Conclusions}
In this paper, asymptotic methods were used to calculate a lower
bound on the free distance that grows linearly with constraint
length for several ensembles of unterminated, protograph-based
periodically time varying LDPC convolutional codes. It was shown
that the free distance growth rates of the LDPC convolutional code
ensembles exceed the minimum distance growth rates of the
corresponding LDPC block code ensembles. Further, we observed that
the performance of the LDPC convolutional codes is consistent with
the iterative decoding thresholds of the underlying protographs.
\section*{Acknowledgement}
This work was partially supported by NSF Grants
CCR02-05310 and CCF05-15012 and NASA Grants NNG05GH736 and
NNX07AK536. In addition, the authors acknowledge the support of the
Scottish Funding Council for the Joint Research Institute with the
Heriot-Watt University, which is a part of the Edinburgh Research
Partnership. Mr. Mitchell acknowledges the Royal Society of
Edinburgh for the award of the John Moyes Lessells Travel
Scholarship.

\end{document}